\documentclass{cs20proc}

\editors{S. J. Wolk}
\publisher{Zenodo}
\conference{The 20th Cambridge Workshop on Cool Stars, Stellar Systems, and the Sun}
\conferencedate{2018}

\title{Exploring the Role of a Tachocline in M-Dwarf Magnetism}

\author{Connor Bice$^{1}$ and Juri Toomre$^1$}
\affiliation{$^{1}$ JILA and Department of Astrophysical and Planetary Sciences, University of Colorado at Boulder, Boulder, Colorado 80309-0440}

\shorttitle{M-Dwarf Tachocline Simulations}
\shortauthors{Bice \& Toomre}

\abs{M-type stars are quickly stepping into the forefront as some of the best candidates in searches for habitable Earth-like exoplanets, and yet many M-dwarfs exhibit extraordinary flaring events which would bombard otherwise habitable planets with ionizing radiation. In recent years, observers have found that the fraction of M-stars demonstrating significant magnetic activity transitions sharply from roughly $10\%$ for main-sequence stars earlier (more massive) than spectral type M3.5 (0.35 M$_\odot$) to nearly $90\%$ for stars later than M3.5. Suggestively, it is also later than M3.5 at which main-sequence stars become fully convective, and may no longer contain a tachocline. Using the spherical 3D MHD simulation code Rayleigh, we compare the peak field strengths, topologies, and time dependencies of convective dynamos generated within a quickly rotating (2 $\Omega_\odot$) M2 (0.4 M$_\odot$) star, with the computational domain either terminating at the base of the convection zone or including the tachocline. We find that while both models generate strong ($\sim$10kG), wreathlike toroidal fields exhibiting polarity reversals, the tachocline model provided a further reservoir for the toroidal field, which slowed the average reversal period from 100 rotations to more than 220 rotations and increased the spectral power of the low-order modes of the near-surface radial field by a factor of 4.}

\begin{document}

\maketitle

\section{Introduction}
M-dwarfs are quickly stepping into the forefront as some of the best candidates in modern searches for habitable, Earth-like exoplanets. This is due mainly to their small masses and luminosities, favoring close-in Goldilocks zones which translate to stronger and more frequent signals for many exoplanet detection schemes. The Goldilocks zone may not provide the whole picture for habitability, however, for many M-dwarfs exhibit extraordinary flaring events \citep{flares} which may bombard these exoplanets with ionizing radiation. 

As flares are primarily magnetic phenomenae, one of the fundamental questions for assessing their effects on exoplanet habitability is "How are they formed?" It has been clear for some time that the magnetic activity a star is capable of generating is closely tied to its rotation rate, with faster rotating stars being more active up to a saturation threshold at roughly $Ro=t_{rot}/t_{conv}=0.1$ \citep{rotact}. Rotation rate alone, however, cannot provide the entire picture.
\subsection{The Tachocline Divide}
Considering the activity of late type stars, a sharp transition can be seen at roughly M3.5 ($0.35 M_\odot$) \citep{tachodiv}. Earlier than M3.5, stars are dominantly inactive with only $10\%$ demonstrating significant markers for magnetism. Among later stars than M3.5, however, nearly $90\%$ display magnetic activity. Suggestively, stellar modeling tells us that it is later than M3.5 where main-squence stars become fully convective (FC). 

Among other things, becoming FC means losing the transition region between the convection zone (CZ) and the underlying radiative zone (RZ). Helioseismology tells us that within the Sun, this transition is a layer of substantial velocity shear and thus it has come to be called the tachocline \citep{soimdi}. The stably-stratified shearing flows of the solar tachocline are often considered to be fundamental in organizing the Sun's dynamo \citep{babcock}. We seek to understand here how their presence or absence may be contributing to the divide observed between early and late type M-dwarfs.

\subsection{Past Dynamo Simulations}
While the tachocline was thought to be critical to the solar dynamo, simulations have shown that a solar-like CZ can, if rotating rapidly enough, sustain globally organized and periodically cycling \it wreaths \rm of magnetism even in the absence of a tachocline. These wreaths can be statistically steady solutions \citep{brownsteady} or go through periodic cycles \citep{browncycle}, among a variety of other behaviors stemming from an intricate and nonlinear parameter space. In the realm of M-dwarfs, several simulations of FC stars, e.g. \citet{browning08} and \citet{yadav15}, have demonstrated very strong magnetism reaching mean toroidal field strengths in excess of 10kG. The simulations with particularly strong fields appear to damp away nearly all of the differential rotation achieved by their hydrodynamic precursors. In some cases, FC simulations have produced broad dipolar caps of magnetism powerful enough to partly suppress convection and create polar dark spots. 

\section{Framing the Problem}
This work employs the open-source 3D MHD code Rayleigh \citep{rayleigh} to evolve the anelastic equations in rotating spherical shells. Rayleigh performs competitively in benchmarks relative to codes such as MagIC and ASH and demonstrates efficient parallelization up to $O(10^5)$ cores. Rayleigh is a pseudospectral code, employing both a physical grid and a basis of spherical harmonics and Chebyshev polynomials. Time stepping is achieved with a hybrid implicit-explicit approach, where the linear terms are advanced via a 2nd order Crank-Nicolson method and the nonlinear terms by 2nd order Adams-Bashforth.
\subsection{The Anelastic Equations}
The anelastic equations are a fully nonlinear form of the fluid equations from which sound waves have been filtered out. This provides an appropriate framework for exploring subsonic convection within stellar interiors, where fast-moving p-modes would otherwise throttle the maximum timestep. The thermodynamic variables are linearized against a one dimensional, time independent background state given by $\bar{\rho},\,\bar{P},\,\bar{T},$ and $\bar{S}$, with deviations from the background written without overbars. The exact form of the anelastic equations solved in Rayleigh is

\begin{equation}
	\nabla\cdot(\bar{\rho}\mathbf{v})=0\;,
\end{equation}
\begin{equation}
\bar{\rho}\frac{D\mathbf{v}}{Dt}=-\bar{\rho}\nabla\frac{P}{\bar{\rho}}-\frac{\bar{\rho}S}{c_p}\mathbf{g}+\nabla\cdot\mathcal{D}\;,
\end{equation}
\begin{eqnarray}
\lefteqn{\bar{\rho}\bar{T}\frac{DS}{Dt}=\nabla\cdot[\kappa\bar{\rho}\bar{T}\nabla S]\;+}\nonumber \\ 
&& 2\bar{\rho}\nu\times[e_{ij}e_{ij}-\frac{1}{3}(\nabla\cdot\mathbf{v})^2]+Q\;,
\end{eqnarray}
\begin{equation}
\frac{\rho}{\bar{\rho}}=\frac{P}{\bar{P}}-\frac{T}{\bar{T}}=\frac{P}{\gamma\bar{P}}-\frac{S}{c_p}\;,
\end{equation}
where $Q$ is the volumetric heating function, $e_{ij}$ is the strain rate tensor, and $\mathcal{D}$ is the viscous stress tensor defined as
\begin{equation}
\mathcal{D}_{ij}=2\bar{\rho}\nu[e_{ij}-\frac{1}{3}(\nabla\cdot\mathbf{v})\delta_{ij}]\;.
\end{equation} 

Due to the resolutions accessible to modern computing, the viscosity, conductivity, and resistivity we employ are not the molecular values, but rather eddy diffusivities. These values are inflated by many orders of magnitude as a parameterization of the turbulent mixing occurring at sub-grid scales.
\subsection{Modeling an M-Dwarf}
\begin{figure}
	\centering
	\includegraphics[width=1.0\linewidth]{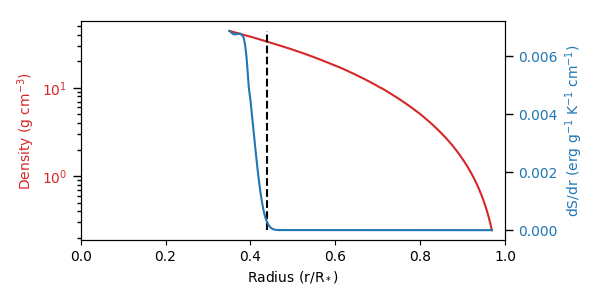}
	\caption{The density stratification (red) and background entropy gradient (blue) employed by the simulations. Due to the numerical noise in MESA's entropy profiles, a smoothing function was applied before taking the gradient which results in a more gradual transition to convective stability than indicated by the stellar model. $R_t$ is marked with a vertical dashed line, while $R_i$ and $R_o$ are at the endpoints of the profiles.}
	\label{fig:reference}
\end{figure}

The calculations were performed within a radial hydrodynamic background state derived using the stellar evolution community code MESA \citep{MESA}. We consider a ZAMS star of 0.4 M$_\odot$ with solar metalicity and rotating at 2 $\Omega_\odot=828\,\mathrm{nHz}$. In the outermost layers of stars, the anelastic equations begin to break down as flows approach the sound speed and non-diffusive radiative transfer becomes important. As a result, we must restrict our computational domain to exclude this region. In our notation, models are either "H" hydrodynamic or "D" dynamo, followed by the frame rotation rate $\Omega_0$ in multiples of $\Omega_\odot$ and lastly by "t" if the computational domain includes the tachocline. A plot of the density stratification and entropy gradient is presented in Figure \ref{fig:reference}. All simulations (H2, H2t, D2, D2t) had CZs extending from $R_t= 0.44R_*$ to $R_o=0.97R_*$, where $R_*=2.588\times 10^{10}$ cm containing $N_r=192$ radial grid points, and spanning $N_\rho=5$ density scale heights. The tachocline models (H2t, D2t) contained an additional radial domain spanning the tachocline region and underlying stable layer, $R_i=0.35R_*$ to $R_t$ with $N_r=48$. An angular resolution of $N_\theta\times N_\phi = 512 \times 1024$ was chosen for all models.

As used previously, e.g. \citet{brownsteady}, we employ viscosity profiles for models H2 and D2 proportional to $\rho^{-0.5}$ where the viscosity at the top of the domain is chosen to be $\nu_0=6.65\times 10^{11} $ cm$^2 $s$^{-1}$. The conductivity $\kappa$ and resistivity $\eta$ are chosen to yield a Prandtl number P$_\mathrm{r}=\nu/\kappa=1/4$ and magnetic Prandtl number P$_\mathrm{rm}=\nu/\eta=4$ throughout the domain. These choices for diffusivity yield a Rayleigh number 1247 times the empirically determined critical point for this system. Models H2t and D2t have similar structure in the CZ, but with dramatic reductions of diffusive amplitudes in the tachocline to increase the viscous time scale and delay its eventual unravelling, as
\begin{equation}
a = a_t+\frac{a_0(\frac{\rho}{\rho_0})^{-0.5}}{1+\exp{(c(R_t-r)/(R_o-r_i))}}\;.
\end{equation}

Here we choose tachocline diffusivities $a_t=10^{-3}a_0$ and a transition steepness $c=200$. In all simulations, the mean $(l=0)$ entropy field sees a separate, much smaller conductivity $\kappa_0$ which serves to discourage thermal conduction as a means of energy transport in the bulk of the CZ, and consequently forces the convective motions to carry the full luminosity of the star.

As is common practice for simulations such as these, the model was first evolved using purely hydrodynamics. After a steady state was achieved, magnetism was introduced as white-noise perturbations and allowed to self-consistently reshape the flows while growing to its mature amplitudes.
\subsubsection{Boundary Conditions} 
The upper and lower boundary conditions are impenetrable and stress free, 
\begin{equation}
v_r|_{\mathrm{bc}}=\frac{d}{dr}(v_\theta/r)|_{\mathrm{bc}}=\frac{d}{dr}(v_\phi/r)|_{\mathrm{bc}}=0\;.
\end{equation}

The lower boundary is thermally insulating, and the top boundary extracts the star's luminosity $L_*=9.478\times 10^{31}$ erg s$^{-1}$ through a fixed conductive gradient, with
\begin{equation}
\frac{dS}{dr}|_{\mathrm{bot}}=0,\; \frac{dS}{dr}|_{\mathrm{top}}=\mathrm{const}\;.
\end{equation}

With no conductive input, energy balance is instead maintained through the volumetric heating function $Q$ which is adapted from the $\epsilon_{nuc}$ and $\nabla\cdot\mathcal{F}_{rad}$ reported by MESA. Finally, the magnetic field matches onto an external potential field at both boundaries, as
\begin{equation}
B=\nabla\Phi,\; \nabla^2\Phi|_{R_i,R_o}=0\;.
\end{equation}

\section{Toroidal Fields}
\begin{figure}
	\centering
	\includegraphics[width=0.85\linewidth]{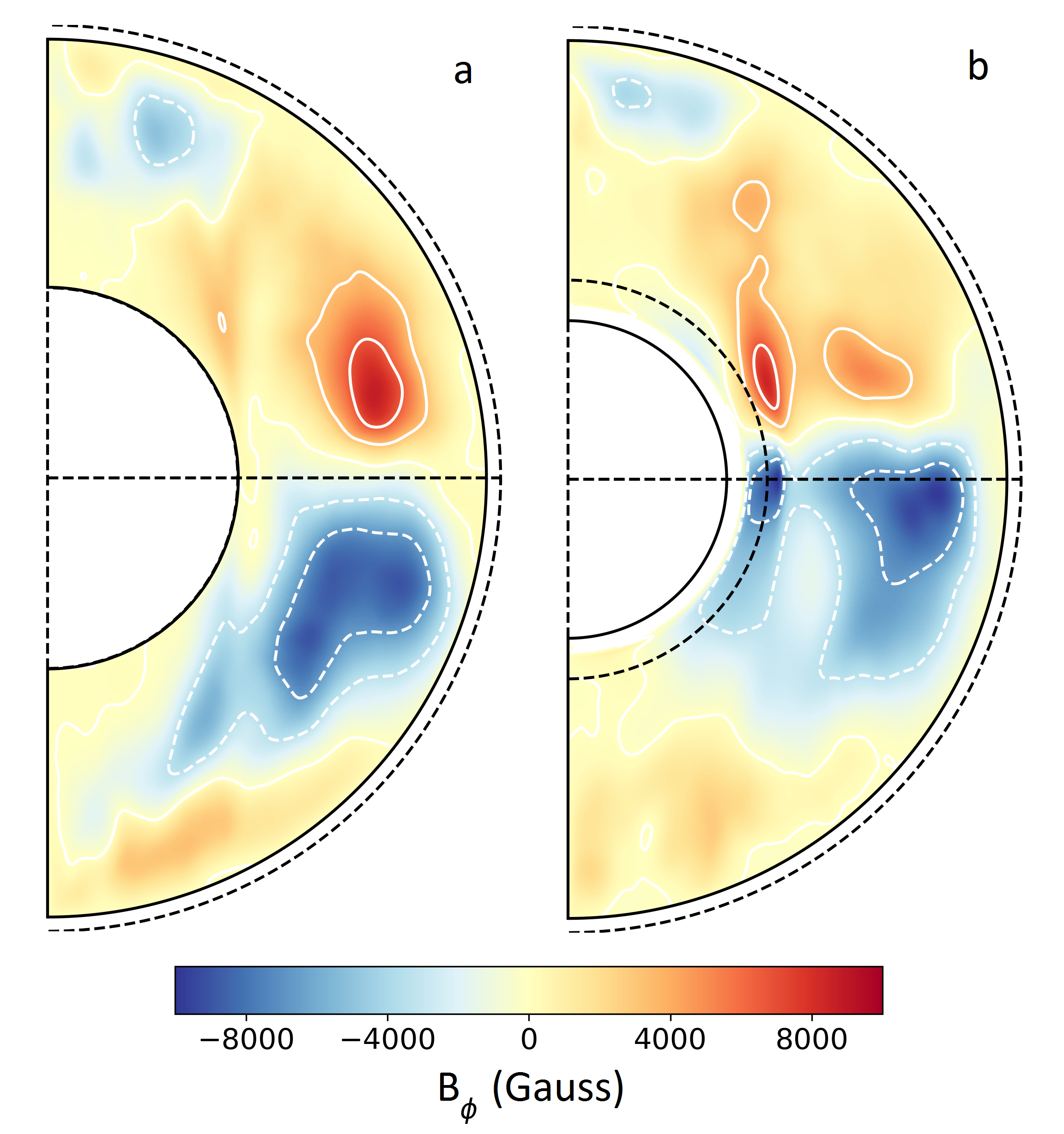}
	\caption{(a) Time- and longitude-averaged $B_\phi$ for case D2, showing the strong wreaths with core averages of 10kG, and the high-latitude structures with average strengths of 2-4kG. (b) The same for case D2t, with a tachocline. In the mid-CZ, the lower hemisphere wreath is dominating and extending across the equator. A second pair of wreaths exist in the tachocline with the same average strengths and polarities as their mid-CZ counterparts at this time, though this parity is not always present.}
	\label{fig:toroidal}
\end{figure}

\begin{figure}
	\centering
	\includegraphics[width=0.85\linewidth]{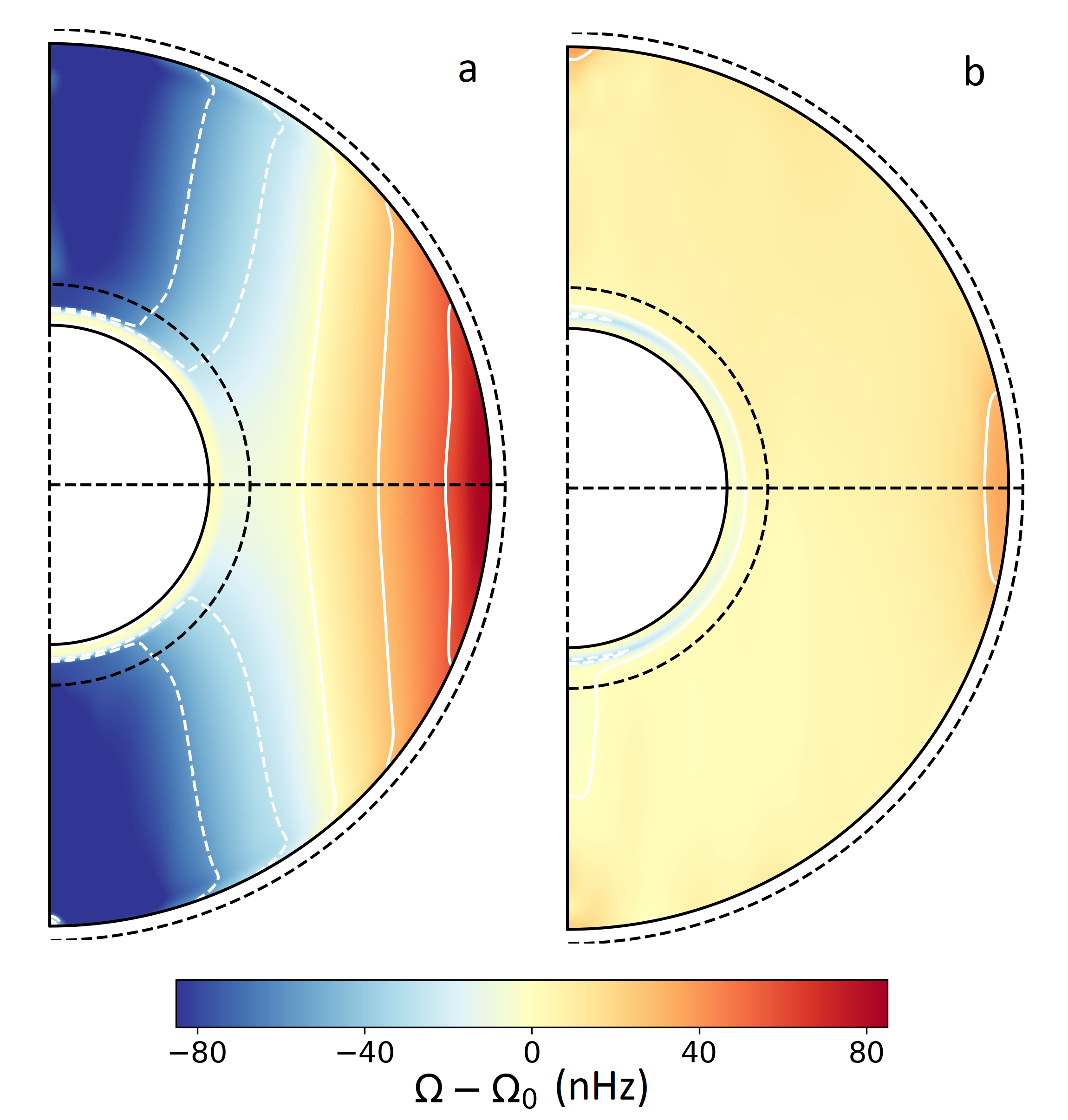}
	\caption{(a) Time- and longitude-averaged $\Omega-\Omega_0$ for case H2t, showing a rotational contrast of 170 nHz from equator to pole in the CZ and a transition to solid body rotation in the RZ. (b) The same for case D2t, where the differential rotation has been all but eliminated in the CZ. A degree of radial shear persists in the tachocline, especially at higher latitudes, as well as in the equatorial near-surface layers.}
	\label{fig:difrot}
\end{figure}

\begin{figure*}[ht]
	\centering
	\includegraphics[width=0.85\linewidth]{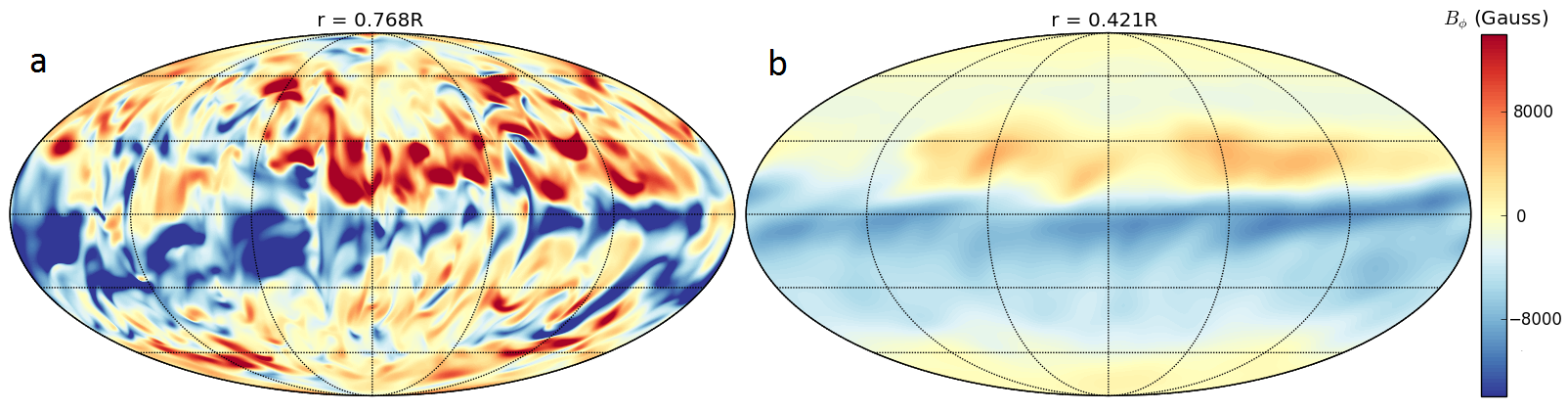}
	\caption{Toroidal fields of model D2t seen in Mollweide projection at mid-depth (a) and in the tachocline (b). Large-scale structures of opposite polarity form in opposite hemispheres, but at the time shown, $t = 1328.5$ rotations, the southern wreath is dominating. The small-scale structures clearly evident in (a) are absent in (b).}
	\label{fig:tslice}
\end{figure*}

\begin{figure*}[ht]
	\centering
	\includegraphics[width=0.95\linewidth]{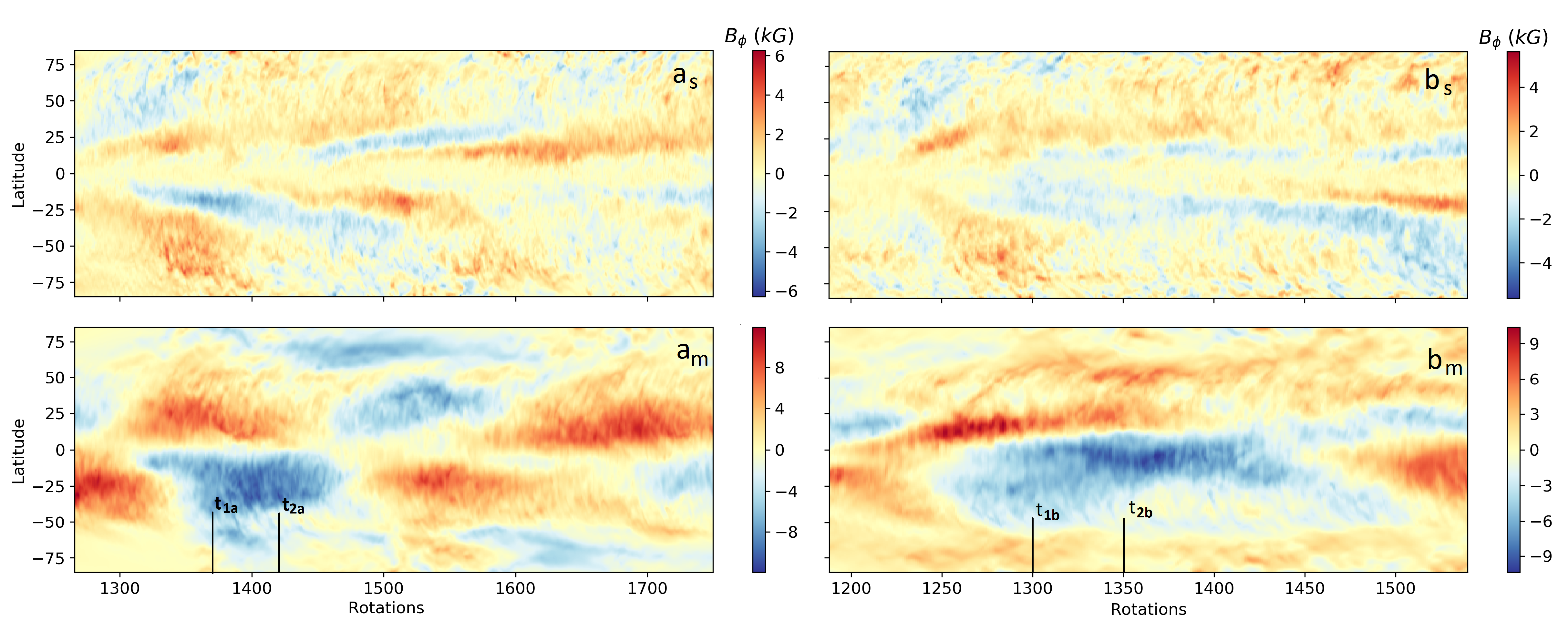}
	\caption{(a$_\mathrm{s}$ and a$_\mathrm{m}$) Azimuthally averaged $B_\phi$ at both depths $r=0.945R_*$ (near surface) and $r=0.684R_*$ (mid-layer), for case D2 as varying in time and latitude. Frequent polarity reversals are evident with a cycle period varying between 100 and 150 rotations, with an outlier at 200 rotations. (b$_\mathrm{s}$ and b$_\mathrm{m}$) The same for case D2t, which underwent only two reversals separated by 220 rotations. Time averaging intervals for Figures \ref{fig:toroidal} and \ref{fig:poloidal} are shown for each model.}
	\label{fig:evolution}
\end{figure*}
Several differences emerge from the two simulations when considering the final configurations of their toroidal fields. As clearly visible in Figure \ref{fig:toroidal}a, model D2 possesses two wreaths of opposite polarity forming at a depth of roughly $0.76R_*$ and at latitudes of about $\pm 25^\circ$. From shell slices as shown for D2t in Figure \ref{fig:tslice}, we see that these wreaths contain substantial longitudinal modulation, most likely an imprint of the vigorous convection taking place in this central region of the CZ. Toroidal field strengths within the wreaths peak close to 20kG, whereas temporal and azimuthal averages hover around 10kG in the wreath cores. Additionally, we see intermittent coherent toroidal fields near the poles. These high-latitude fields are weaker than those in the wreaths, achieving peak strengths of 16kG and azimuthal averages of about 4kG.

The inclusion of a tachocline in model D2t alters the character of the mid-CZ wreaths. While the field strengths and choppiness remain unaffected, we note from Figure \ref{fig:tslice} a tendency for one wreath to dominate over the other in magnitude. This in turn allows the dominant wreath to push closer to the equator and in some cases extend a few degrees across it. The same intermittent behavior at high latitudes as in case D2 is observed here, though with marginally weaker amplitudes.

We find that the tachocline provides a reservoir for the toroidal field as can be seen in Figure \ref{fig:toroidal}b. The wreaths here are for the most part not produced locally, and are instead pumped into this region of reduced resistivity by overshooting convection. In solar models, the shear of the tachocline is thought to provide a mean field $\Omega$-effect for converting poloidal to toroidal field, but that is not occurring in this simulation. As evident in Figure \ref{fig:difrot}, the strong fields have damped the latitudinal contrast in rotation rate from 170 nHz ($21\%$) in H2t to just 30 nHz ($3.6\%$) in D2t. Since the CZ has been brought nearly to solid body rotation, the transition to RZ mandates less rotational shear. In a different region of parameter space, with weaker fields through slowed rotation or reduced P$_\mathrm{rm}$, for example, we might expect to find that some differential rotation remains in the CZ and thus the tachocline could contribute a more significant $\Omega$-effect for toroidal field generation.

With less disruption by the turbulent convective motions, we find that the wreaths in the tachocline shown in Figure \ref{fig:tslice}b are nearly uniform in longitude and thus their peak field strengths are close to their average core strengths at 10kG.

\subsection{Time Evolution}
In examining the evolution of the toroidal fields over time, we find further differences in the magnetic behaviors of these two simulations. Figure \ref{fig:evolution} shows that both models undergo reversals in the polarities of their mean fields, though only case D2 exhibits any regularity in its cycling period during the time captured by the simulations here. D2 has its toroidal fields reverse in both hemispheres every 100 to 150 rotations, with a failed reversal in the southern hemisphere near rotation 1610 leading to a half-cycle of twice the usual length in both hemispheres and a period where both wreaths had the same polarity. 

With a tachocline, however, we observe only two reversals in the 1550 rotations captured so far in simulation D2t. These reversals are separated by roughly 220 rotations, longer than any half-cycle in the CZ-only simulation. The latter of the two reversals occurred just before the end of the run for case D2t, so it is not clear yet whether these reversals and thus the interval between were random occurrences or if they mark the onset of a cycling phase for this star's dynamo.

\section{Poloidal Fields}
In addition to its effects on the internal toroidal fields, inserting a tachocline at the base of our CZ has led to significantly stronger and more organized poloidal fields near the stellar surface. We consider the total spectral power of the radial field at the top of the domain $S=\sum_{l,m}f^2_{lm}$ where $f_{lm}$ is a spherical harmonic coefficient. With no stable layer, we find $S=5.68$ kG$^2$, while case D2t has a surface spectral power of 6.63 kG$^2$, an increase of $17\%$.

\begin{figure}
	\centering
	\includegraphics[width=0.85\linewidth]{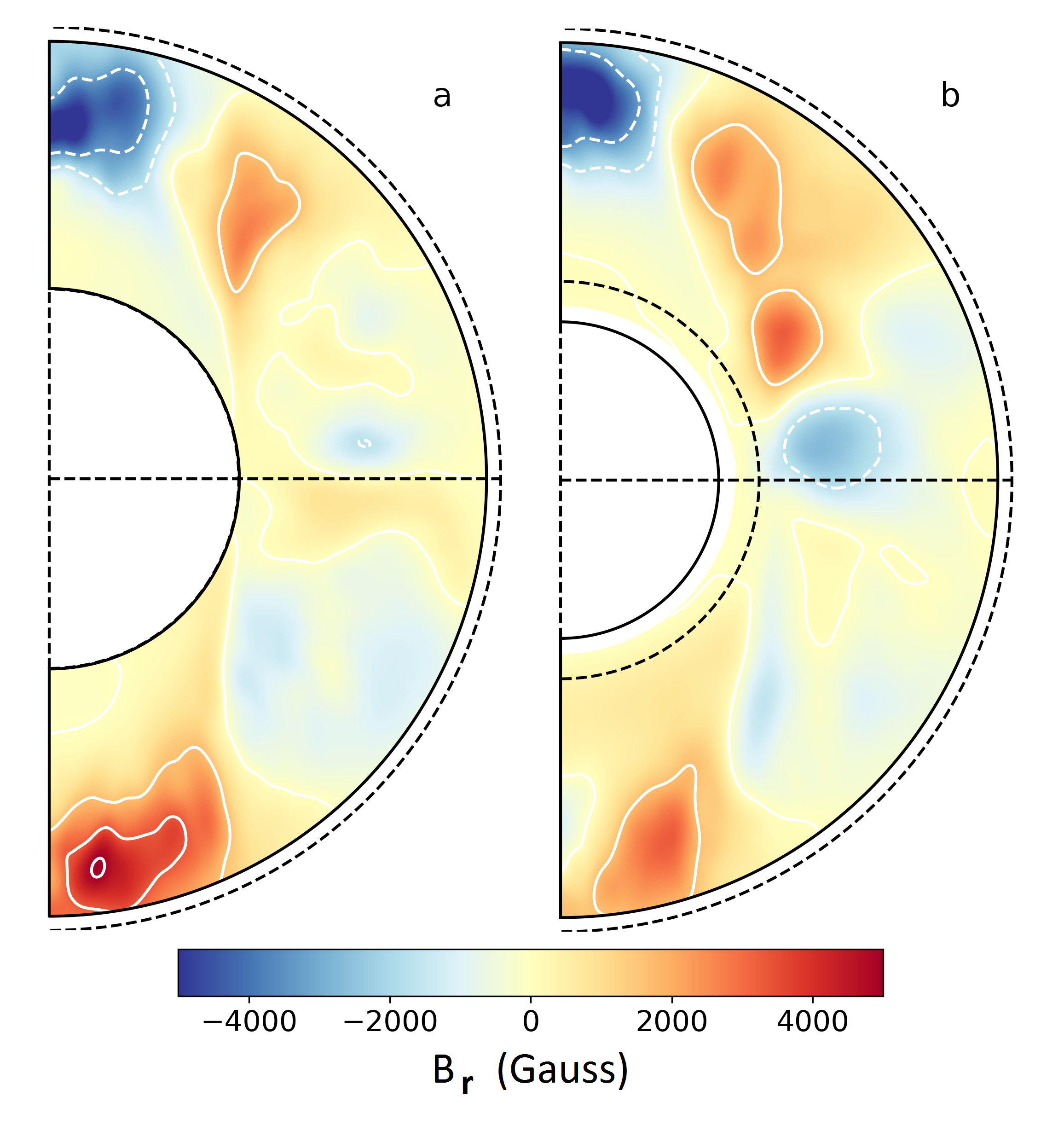}
	\caption{(a) Time- and longitude-averaged $B_r$ for case D2, showing high latitude caps of poloidal field at strengths in excess of 5kG along with weaker mid-latitude fields. (b) The same for case D2t, low and mid-latitude structures are similar to case D2 but with greater amplitude, and the polar caps show more asymmetry between the two hemispheres.}
	\label{fig:poloidal}
\end{figure}

\begin{figure}
	\centering
	\includegraphics[width=1.0\linewidth]{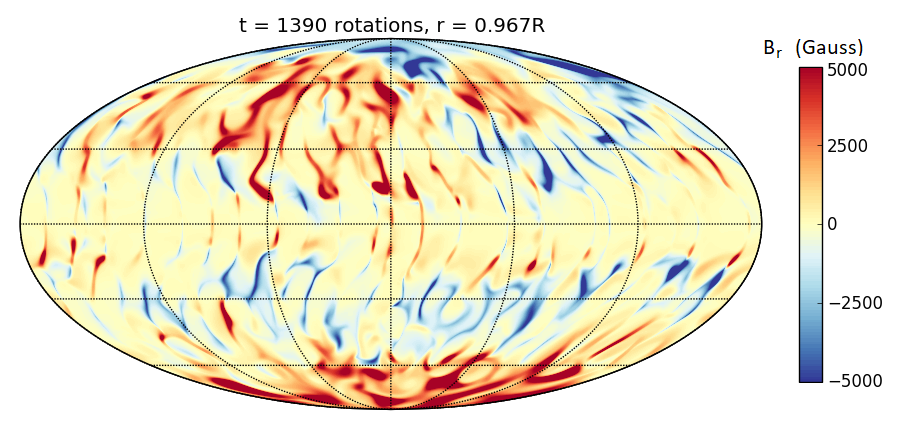}
	\caption{Radial fields of model D2 shown in Mollweide projection near the upper boundary of the simulation at depth 0.967R$_*$. Near the equator, the fields trace downflow lanes and reach magnitudes of 2-6 kG, while high-latitude fields have greater filling factors and attain strengths on the order of 10kG.}
	\label{fig:pslice}
\end{figure}

Figure \ref{fig:pslice} shows that the character of the surface magnetism varies significantly with latitude. Near the equator, the radial fields trace the narrow convective downflow lanes. In some instances, the concentrated fields become intense enough to fully suppress their local convection. There is an analogy to be made with starspots here, but we must keep in mind that these simulations extend only to $0.97R_*$, and thus such structures would still need to extend through the more turbulent layers of the star before they could be observable. At high latitudes, in the shadow of the tangent cylinder, the radial field covers much more area and typically has stronger fields.

The structure present at high-latitude is suggestive of polar caps, and indeed an azimuthal average as shown in Figure \ref{fig:poloidal} confirms their presence. Spherical harmonic decomposition reveals another role played by the tachocline: organizing surface poloidal fields into large-scale structures. Not only does case D2t have more power in its poloidal field at the surface, but the fraction which is axisymmetric $S_{AS}=\sum_{l}f^2_{l,m=0}$ is nearly twice as great at 0.306 as that of case D2 at 0.158. While neither model produced a dipole dominated field, with axisymmetric dipole fractions $ f_{1,0}^2/S$ of 0.004 and 0.026 for D2 and D2t, case D2t showed a strong preference for its axisymmetric quadrupole and octupole modes, which together contained $(f^2_{2,0}+f^2_{3,0})/S=0.193$ of the power in the radial field. By way of contrast, these modes contained only $0.054$ of the total spectral power in case D2.

\subsection{Spin-Down Implications}
The low-order modes of the surface poloidal field are particularly important due to their interactions with the stellar winds and consequently with the spin-down histories of these stars. Since the radial decay of magnetic multipoles goes as $r^{-(l+1)}$, the effective lever-arm each mode could use to exert a torque shrinks very quickly with increasing $l$. Thus, magnetized wind spin-down analyses tend to focus on modes with $l=1,2,$ or 3 \citep{winds}. 

Applying this principle qualitatively to the field configurations achieved in our models, we see some hints that by enhancing the low order poloidal fields near the surface, a tachocline may cause a star to spin-down more rapidly. We must emphasize that our comparison is not between early and late M-dwarfs, but rather between an early M-dwarf and another model of an early M-dwarf whose tachocline has been replaced with an impenetrable boundary.

\section{Conclusions}
We have compared the dynamos operated and fields generated in simulations of two M2-like stars, differing initially only in that one has a computational domain including only the CZ, while the other also includes a portion of the underlying RZ. In doing so, we have arrived at three main conclusions regarding the magnetism of such stars and how the presence of a tachocline may modify it:
\begin{enumerate}
\item The CZs of early M-dwarfs are perfectly capable of generating and organizing strong toroidal fields with or without an underlying tachocline of shear.
\item The tachocline can provide a reservoir for the fields produced in the bulk of the CZ, and coupling between this reservoir and the mid-CZ dynamo can slow the reversals of the global field.
\item The tachocline helps to organize near-surface poloidal fields onto larger spatial scales, which may create a favorable condition for the host star to shed angular momentum through its magnetized wind.
\end{enumerate}

While few experiments of this type have been conducted for early M-dwarfs, the more extensively studied parameter spaces in the solar regime have proven to house a rich diversity of behavior. More work is currently underway to examine the local sensitivities of our model in parameter space, and thus to assess the robustness of these conclusions concerning the features of deep convective shells with underlying tachoclines.  

\section*{Acknowledgements}
{We thank Ben Brown, Sacha Brun, and Brad Hindman for helpful advice in developing this work. We thank Nick Featherstone for his assistance with Rayleigh, as well as the Computational Infrastructure for Geodynamics (http://geodynamics.org) which is funded by the National Science Foundation. The calculations presented here were performed on the NASA Pleiades supercomputer. This work was supported by NASA grant NNX17AG22G.}

\bibliography{CS20}
\bibliographystyle{cs20proc}

\end{document}